\documentclass[final,authoryear,3p,times,twocolumn]{elsarticle}
\usepackage{graphicx}
\usepackage{epsfig}
\usepackage{amssymb}

\journal{New Astronomy}
\begin{document}
\begin{frontmatter}
\title{500 days of Stromgren b, y and narrow-band [OIII], Ha photometric evolution of gamma-ray Nova Del 2013 (= V339 Del)}

\author[un,um]{Ulisse Munari}
\author[af]{Alessandro Maitan}
\author[af]{Stefano Moretti}
\author[af]{and Salvatore Tomaselli}

\address[un]{corresponding author: Tel.: +39-0424-600033, Fax.:+39-0424-600023, e-mail: ulisse.munari@oapd.inaf.it}
\address[um]{INAF Astronomical Observatory of Padova, via dell'Osservatorio 8, 36012 Asiago (VI), Italy}
\address[af]{ANS Collaboration, c/o Astronomical Observatory, 36012 Asiago (VI), Italy}

\begin{abstract}

We present and discuss highly accurate photometry obtained through medium
Stromgren $y$, $b$ bands and narrow [OIII], H$\alpha$ bands covering 500
days of the evolution of Nova Del 2013 since its maximum brightness.  This
is by far the most complete study of any nova observed in such photometric
systems.  The nova behaviour in these photometric bands is very different
from that observed with the more conventional broad bands like
$U$$B$$V$$R_{\rm C}$$I_{\rm C}$ or $u^\prime g^\prime r^\prime i^\prime
z^\prime$, providing unique information about extent and ionization of the
ejecta, the onset of critical phases like the transition between optically
thick and thin conditions, and re-ionization by the central super-soft X-ray
source.  The actual transmission profiles of the $y$, $b$, [OIII] and
H$\alpha$ photometric filters have been accurately measured at different
epochs and different illumination angles, to evaluate in detail their
performance under exact operating conditions.  The extreme smoothness of
both the H$\alpha$ and [OIII] lightcurves argues for absence of large and
abrupt discontinuities in the ejecta of Nova Del 2013.  Should they exist,
glitches in the lightcurves would have appeared when the ionization and/or
recombination fronts overtook them.  During the period of recorded very
large variability (up to 100$\times$ over a single day) in the super-soft
X-ray emission (from day +69 to +86 past $V$ maximum), no glitch in excess
of 1\% was observed in the optical photometry, either in the continuum
(Stromgren $y$) or in the lines ([OIII] and H$\alpha$ filters), or in a
combination of the two (Stromgren $b$, Johnson $B$ and $V$).  Considering
that the recombination time scale in the ejecta was one week at that time,
this excludes global changes of the white dwarf as the source of the X-ray
variability and supports instead clumpy ejecta passing through the line of
sight to us as its origin.

\end{abstract} 

\end{frontmatter}

\section{Introduction}
\label{}

The vast majority of the photometric observations carried out on novae at
optical wavelengths has been obtained with Johnson {\it UBV} and Cousins
$R_{\rm C}$$I_{\rm C}$ photometric systems.  The introduction of $UBVR_{\rm
C}I_{\rm C}$ equatorial standards by Landolt (1973, 1983, 1992, 2009),
recently expanded to +50$^\circ$ declinations (Landolt 2013), together with
the almost exclusive use of CCDs as detectors during the last 25 years, has
brought more order and uniformity to the photometry of novae.  In recent
years, the Sloan $u^\prime g^\prime r^\prime i^\prime z^\prime$ system
(Fukugita et al.  1996, Smith et al.  2002) is being progressively
introduced into leading observing facilities, and the amount of photometry
of novae collected with this system is slowly increasing.

Common to both the Johnson-Cousins and Sloan systems is the large width of
their photometric bands, that trades diagnostic capability with the appeal
to reach fainter magnitudes.  Novae have very complex spectra, dominated
by continuum emission around maximum brightness and by a few very strong
nebular lines during advanced decline.  As a consequence, the astrophysical
meaning of photometry collected in the Johnson-Cousins and Sloan systems
steadily declines in pace with the decline of a nova from optical maximum, as
the percentage of the recorded flux shifts more and more away from continuum
and toward a few emission lines.  These emission lines may lie close to the
edges of the photometric bands (expecially true for [OIII] 4959, 5007 \AA\
and the $B$ and $V$ bands), where small differences from one set of filters
to another may spoil completely the possibility to meaningfully combine data
on the same nova collected with different instruments.

Optical photometry of novae in photometric systems other than $UBVR_{\rm
C}I_{\rm C}$ is exceedingly rare.  This is not due to the paucity of
photometric systems - more than 200 of them are documented in the Asiago
Database of Photometric Systems (Moro and Munari 2000, Fiorucci and Munari
2003) - but to the unpredictability of nova eruptions.  Thus, when a nova
erupts, an observer rushes to the telescope (or places an urgent request for
service observing) and uses whatever photometric device is already mounted
and available, usually equipped with the general purpose $UBVR_{\rm C}I_{\rm
C}$ or $u^\prime g^\prime r^\prime i^\prime z^\prime$ filters.  This is even more
the case at later times, when observations of the fading nova requires longer
integrations and are more diluted in time.

Nonetheless, some timid attempts to observe novae in photometric systems
other than $UBVR_{\rm C}I_{\rm C}$ have been carried out in the past.  The
second most used photometric system for general astronomical application has
been the {\it uvby} Stromgren (1956) system (cf the General Catalog of
Photometric Data\footnote{http://obswww.unige.ch/gcpd/gcpd.html} by
Mermilliod, Mermilliod and Hauck 1997).  At the time of single channel
photo-electric photometry, a few novae have been observed in this system,
notably Nova Cyg 1975 (Lockwood and Millis 1976) and Nova Cyg 1978
(Gallagher et al.  1980, Kaler 1986).  These pionering attempts soon
recognized the value of the Stromgren $y$ band in tracing the emission in the
continuum without contamination by emission lines.  However, these early efforts did
not continue into the CCD era, presumably because of the
difficulties of manufacturing Stromgren interference filters in larger
formats and using them in {\em converging} beams instead of the {\em
parallel} beams of earlier photo-electric photometers.

In recent years, classical interference filters has been replaced by far
more efficient multi-layer dielectric filters, removing the obstacles in
using them with large CCD detectors and fast converging light beams. 
Consequently, it seems now appropriate to reconsider the Stromgren {\it
uvby} filters in the context of diagnostic photometry of nova outbursts.  In
this paper we present our extensive Stromgren {\em b,y} photometry of Nova
Del 2013, densely and accurately covering the first 500 days of its
outburst, in what appears to be - to the best of our knowledge - the best
ever recorded Stromgren lightcurve for a nova.  

The photometry of novae through a narrow filter isolating a single
emission line has been even less popular in the past, one such attempt being the
photometry of Nova Del 1967 through an H$\alpha$ filter by Mannery (1970). 
In parallel with Stromgren {\em b,y} filters, we have also followed the
evolution of Nova Del 2013 through narrow filters transmitting H$\alpha$ and
[OIII] 5007 \AA\ lines.  This paper represents the first actual
implementation of the Stromgren/Narrow-band photometry we initially explored
on Nova Mon 2012 (Munari et al.  2013) by integrating the filter
transmission profiles on accurately fluxed spectra. A comprehensive
discussion of the performances on novae of Stromgren and narrow band filters 
is postponed to when at least a few additional other novae have been observed
through these filters.

\section{Overview of Nova Del 2013}

Nova Del 2013 was discovered at $V$=6.8 on Aug 14.584 UT by K. Itagaki
(Nakano et al.  2013), and was designated PNV J20233073+2056041 when it was
posted to the TOCP webpage.  Spectroscopic classification as a FeII-class
nova was provided by Darnley et al.  (2013) from Aug 14.909 UT spectra and
by Shore et al.  (2013a) from Aug 14.87 UT spectra. 

The nova progenitor is the faint $B$=17.2 star USNO B-1 1107-0509795.
Denisenko et al. (2013) reported the nova still in quiescence at $\sim$17.1
mag fourteen hours before the discovery by Itakagi, which indicates a very
fast rise to maximum. This well agrees with the modeling of CHARA Array
infrared interferometric observations by Schaefer et al. (2014) that fix
around Aug 14.3 the start of the observed fireball expansion of Nova Del
2013. Deacon et al. (2014) searched the PanSTARRS-1 image archive and
identified twenty-four observations in the 1.2 years before outburst, during
which the progenitor did not brighten significantly in comparison with the
archival plate photometry reported by Munari and Henden (2013). The latter found
the mean brightness of the progenitor, as recorded on Asiago plates exposed
between 1979-1982, to be $B$=17.27 and $V$=17.60, for a mean color index
$(B-V)$=$-$0.33, and measured a mean $B$=17.33 from APASS observations in
2012.  The recorded total amplitude of variation in B band is 0.9 mag, with
color and amplitude in agreement with a progenitor dominated by the emission
from an accretion disc. Such values are in excellent agreement with USNO-B1
values from Palomar surveys 1 and 2 (from plates exposed in 1951 and 1990),
arguing for a long term stability of the progenitor before the 2013
eruption.

Various preliminary reports were made on early optical photometric evolution
of Nova Del 2013 (CBET 3628, CBET 3634, IAUC 9258, Tomov et al.  2013,
Munari et al.  2013b, Chochol et al.  2014).  Munari et al.  (2013d)
presented and discussed an accurate and densely mapped $B$,$V$,$R_{\rm
C}$,$I_{\rm C}$ lightcurve of Nova Del 2013 covering the time interval from
pre-maximum until after the transition from otically thick to optically thin
ejecta, two months into the decline.  Maximum brightness in $V$ and $B$
bands occurred around August 16.4 UT (JD 2456520.9, the $t_0$ from which time
is counted in this paper) at $V$$\sim$4.46 and $B$$\sim$4.70, thus
setting the outburst amplitude to $\Delta$$B$=12.5 mag.  The characteristic
decline times were $t^{B}_{2}$=12, $t^{B}_{3}$=30 and $t^{V}_{2}$=10.5,
$t^{V}_{3}$=23.5 days, which place Nova Del 2013 in a borderline position
between fast and very fast novae according to the Warner (1995)
classification scheme.  The lightcurve of Munari et al.  (2013d) shows a
smooth decline for Nova Del 2013, a brief plateau appearing soon after
maximum brightness and lasting ∼2 days, a longer plateau extending from
about day +20 to +37 that started when the nova had declined by
$\Delta$$V$=2.8 mag from maximum brightness, the transition around day +62
from optically thick to optically thin conditions of the ejecta occurring
$\Delta$$V$$\sim$6.0 mag down from maximum brightness, and blue intrinsic
colors $(B-V)$=+0.14 at maximum and $(B-V)$=+0.04 at $t^{V}_{2}$, that
become $(B-V)_\circ$=$−$0.04 and $(B-V)_\circ$=$−$0.14 once corrected for
the low $E_{B-V}$=0.18 insterstellar reddening affecting the nova (cf. 
Tomov et al.  2013, Munari et al.  2013a).

A rich ensamble of preliminary reports described the early evolution of the
optical and UV spectrum of Nova Del 2013 (Shore et al. 2013a,b,c,d,e,f,
Shore et al. 2014, Munari et al. 2013a,b,c, Darnley et al. 2013, Darnley and
Bode 2013, Tomov et al. 2013, Tarasova 2013, Tarasova and Shakhovskoi 2013,
Woodward et al. 2013, Skopal et al. 2014a), with a more structured report
presented by Skopal et al. (2014b). The wealth of optical spectroscopic
information so far circulated is so large that is practically impossible to
even quickly summarize it, not to mention the fact that huge spectroscopic 
databases on Nova Del 2013 (including ours, to be studied elsewhere) have not 
yet been studied. Here it would suffice to say that the spectral
evolution of Nova Del 2013 has followed the path of typical FeII-class novae,
with expansion velocities ranging from 610 and 2500 km/s (depending on the
spectral feature and epoch, which in turn depend of the optical depth, 3D
structure and ionization of the ejecta that evolve with time). Pre-maximum
spectra showed extremely strong P-Cyg profiles on Balmer and FeII lines,
and pure HeI aborptions. At the time of optical maximum, most of the
emission had gone and the deep absorptions reinforced; only CaII H\&K
and H$\alpha$ remained in prominent emission, of similar intensity and 
flanked by very deep P-Cyg absorptions. The first days into decline saw
the progressive reduction of absorption lines, and re-emergence of emission by Balmer
and FeII lines. Five days past optical maximum the P-Cyg absorptions disappeared
from all lines (except for CaII H\&K), at the time
when FeII lines peaked in their intensity compared to Balmer lines. With
passing time the excitation and ionization of the ejecta increased, with
HeI appearing in emission in September and HeII in October 2013. The
transition from optically thick to thin conditions, marked conventionally by
[OIII] surpassing in strength H$\beta$, occurred in mid October 2013, and
[OIII] became stronger than the blend H$\alpha$+[NII] in early January 2014
marking the transition toward advanced nebular conditions. The profile of
emission lines have shown remarkable changes along the evolution, before
settling on a distinct double peaked shape, with a separation of $\sim$1100
km/sec for HeII and $\sim$900 for [OIII].

Schaefer et al. (2014) performed interferometric observations of Nova Del
2013 with the CHARA Array during the first 43 days of the outburst. They
detected an ellipticity in the light distribution, suggesting a prolate or
bipolar structure that develops as early as the second day. Combining the
angular expansion rate with radial velocity measurements, they derived a a
geometric distance to the nova of 4.5$\pm$0.6 kiloparsecs from the Sun. The
maximum-magnitude-to-rate-of-decline (MMRD) relation (Downes and Duerbeck
2000) gives a distance ranging from 3.3 kpc to 4.1 kpc, depending on the
adoption of the linear or nonlinear MMRD formulations.

Nova Del 2013 has been subject of intensive monitoring by the X-ray Swift
satellite and sparser visits by Chandra and XMM (Kuulkers et al.  2013,
Nelson et al.  2013, Osborne et al.  2013b, Page et al.  2013, Page et al
2014).  First detection of very weak X-ray emission occurred on day +33,
and was consistent with shocked gas in an expanding nova shell with no
evidence for super-soft emission.  The latter began to emerge, albeit at
very low flux levels, on day +58.  It was only on day +69 that the soft
X-rays flux started to increase, reaching 1 count/sec over the 0.3$-$1.0 keV
energy range.  The super-soft X-ray emission reached a peak value of
$\sim$100 counts/sec on day +86.  At the same time a quasi-periodic
oscillation (QPO) of 54 sec was present in the data.  This periodicity was
confirmed by Newton-XMM X-ray observations for day +97.  This QPO is similar
to the one seen at 35 sec in both nova RS Oph and nova KT Eri.  The
origin of this short-period QPO is uncertain, possibly being related to
either the spin of the white dwarf or an oscillation in the nuclear burning
envelope around the white dwarf.  The super-soft phase was still well underway (at
40 counts/sec) at the time of the last Swift visit on day +143 before the
satellite observations were interrupted because of Sun-angle contraints. 
When they resumed on day +200, the soft count rate had dropped to $\sim$0.8
count/sec and further reduced to $\sim$0.4 counts/sec by day +205, meaning
the end of the super-soft phase and therefore of the sutained nuclear
burning at the surface of the white dwarf.

One of the most surprising discoveries of recent years about novae has been
their detection in $\gamma$-rays following the launch of the sensitive Fermi
satellite.  After the initial serendipitous detections at energies $\geq$100
MeV of novae V407 Cyg in 2010 and V959 Mon in 2102, three more bright novae
have been pointed at and detected by Fermi in the $\gamma$-rays: Nova Sco 2012,
Nova Cen 2013, and Nova Del 2013, making the latter a full member of a still
very exclusive club (Abdo et al.  2010, Cheung et al.  2010, 2012, 2013,
Cheung, Glanzman and Hill 2012, Hays et al 2013, Laird et al.  2013).  The
origin of the $\gamma$-rays recorded from these novae is still debated, but
a consensus is growing around the idea that the strong shocks needed to
accelerate particles at relativistic energies develop within a mixed and
delayed torus-bipolar expansion pattern for the ejecta (Ackermann et al. 
2014, Chomiuk et al.  2014, Metzger et al.  2015). A search for TeV emission
from Nova Del 2013 has provided negative results (Sitarek et al. 2015).

Radio observation of Nova Del 2013 have been carried out on a number of
dates (Anderson et al.  2013, Chomiuk et al.  2013a,b, Roy et al.  2013). 
Initial non detections were used to infer a distance to the nova in excess
of 2 kpc.  The first positive detection occurred on day +22.6 at 0.31, 0.82
and 1.06 cm while the nova remained undetected at 4.05 cm and longer
wavelengths, consistent with an optically thick thermal emission.  The nova
was then detected at 2.0 cm on day +40.

A great deal of IR data have been collected on Nova Del 2013 (Banerjee et
al.  2013a,b, Burlak et al.  2013, Cass et al.  2013a,b,c,d,e, Gehrz et al. 
2013, Rudy et al.  2013, Shenavrin et al.  2013, Schaefer et al.  2014,
Taranova et al.  2014).  Starting around day +30 a limited amount of dust
formed in the ejecta of Nova Del 2013.  The dust remained optically thin,
absorbing and re-emitting about 10\% of the total luminosity radiated by the
nova, and cooling from $\sim$1500~K on day +36, to $\sim$1100~K on day +56,
to $\sim$700~K on day +100.  Near-IR spectra were dominated by strong
emission lines, in particoular C I lines which are hallmarks of the
FeII-class of novae (Banerjee and Ashok 2012), and followed the expected
behaviour for a normal nova of this type.  Hints of a triple
peaked structure were seen in the HI line profiles, suggesting a bipolarity
in the flow.

   \begin{table}
     \caption{Our measurement for the transmission profiles of the photometric
     filters used in this study, for normal incidence and scaled to 1.00 at maximum.} 
      \centering
      \includegraphics[width=61mm]{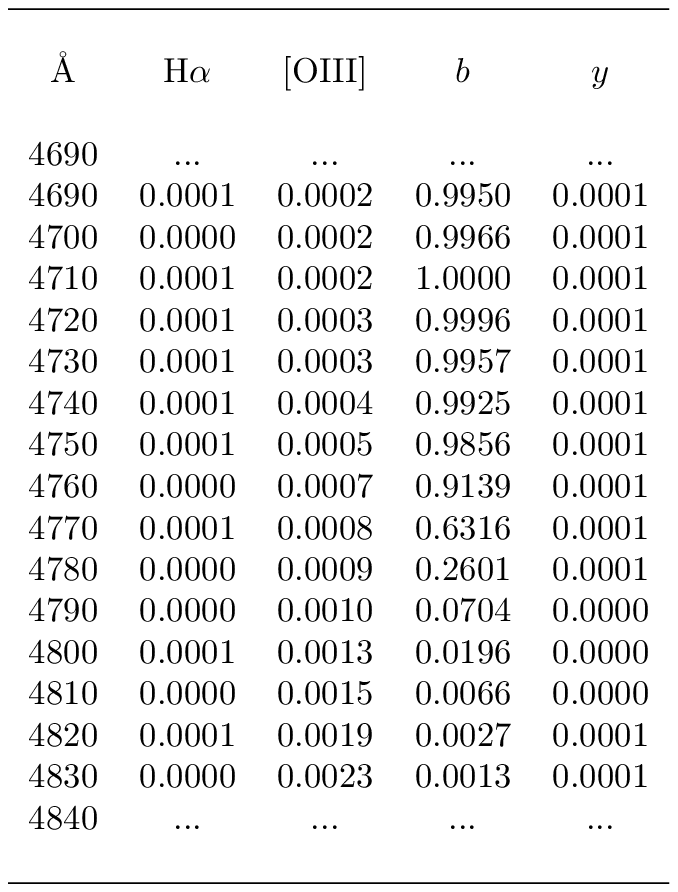}
      \label{tab1}
  \end{table}

 \begin{figure*}[!Ht]
     \centering
     \includegraphics[angle=270,width=15.2cm]{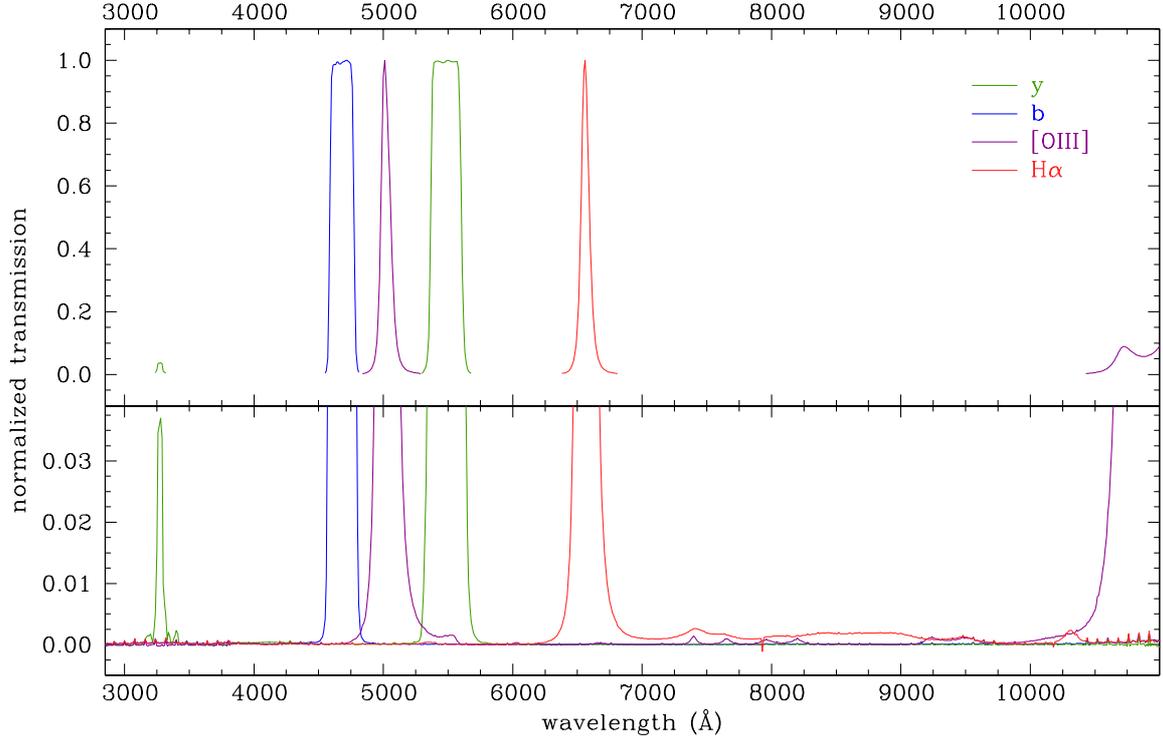}
     \caption{Transmission of the Astrodon Stromgren
     $b$,$y$ and Baader narrow-band [OIII], H$\alpha$ filters we used in this
     study of the photometric evolution of Nova Del 2013.  {\em Top panel:}
     transmission above 0.3\% of the peak is plotted.  {\em Bottom panel:}
     zooming on the residual transmission at the lowest levels.}
     \label{fig1}
  \end{figure*}

   \begin{table*}[!Ht]
     \caption{Summary of basic parameters we measured for the photometric
     filters.  Peak \%: transmission percentage at its peak; $\lambda_{\rm
     eff}$: the photo-center of the transmission profile WHM: width of the
     profile at 50\% of its peak transmission; $\lambda$(10\%),
     $\lambda$(50\%), $\lambda$(90\%): wavelength intervals at which the
     filter transmission reaches 10\%, 50\% and 90\% of its peak value,
     respectively.  All numbers are given for two configurations: normal
     light incidence (90$^\circ$ deg), and average incidence (86$^\circ$.1
     deg) at an f/5.3 Newtonian focus with central obstruction.}
      \centering
      \includegraphics[width=130mm]{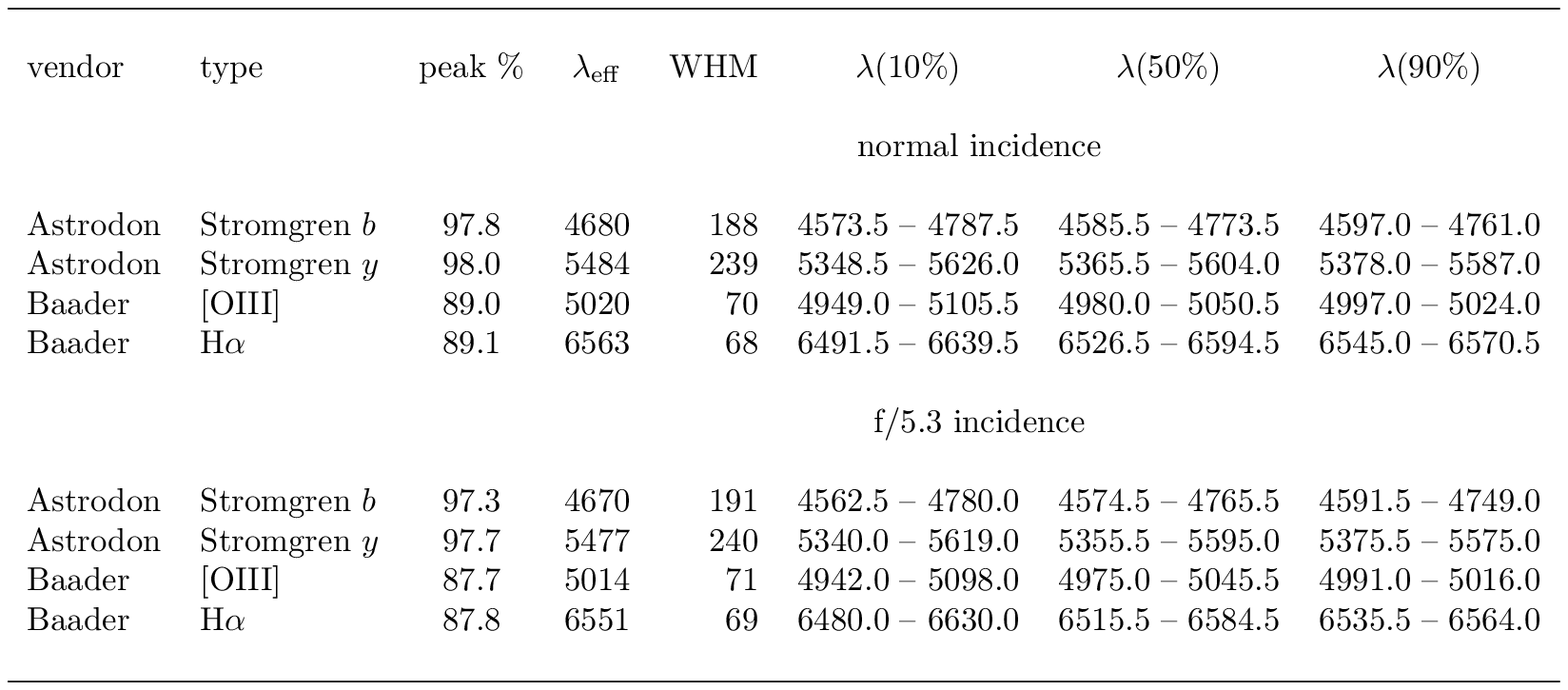}
      \label{tab2}
  \end{table*}

   \begin{figure*}[!Ht]
     \centering
     \includegraphics[width=15.2cm]{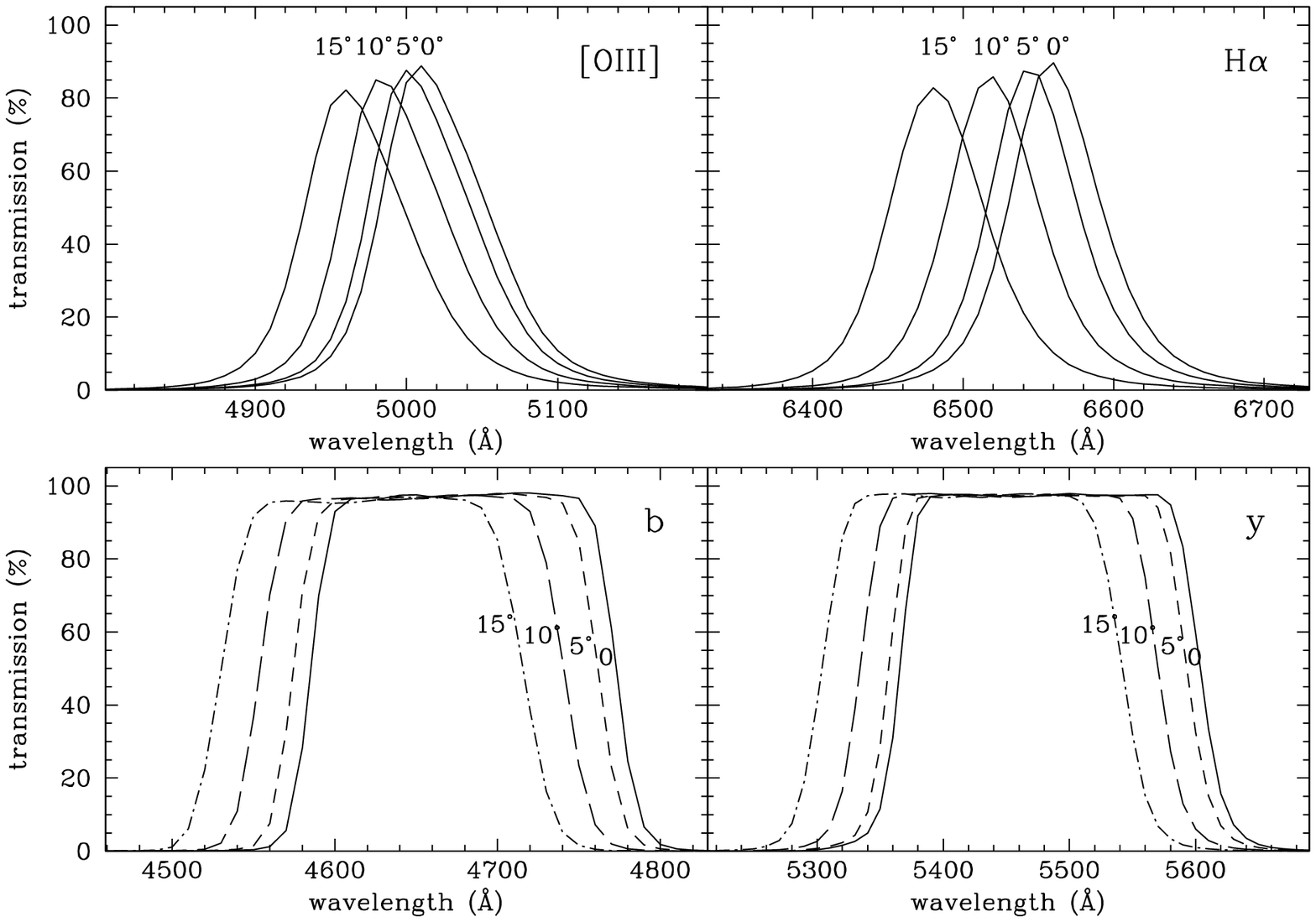}
     \caption{Dependence of the filter transmission profiles on the angle of
     incidence ($\theta$=0, 5, 10 and 20 deg).}
     \label{fig2}
  \end{figure*}

\section{Characterization of filters}

The Stromgren $b$, $y$ filters were purchased from Astrodon and the [OIII]
and H$\alpha$ filters from Baader at the beginning of 2012.  At that
time we measured their transmission and repeated the operation with the same
instrumentation two and half years later in September 2014, to evaluate the
impact of aging and continued use at the telescope.

To measure the transmission profiles, we used a Perkin Elmer UV-Vis
spectrometer Lambda Bio 40 operated by the ARPA laboratory of Forli (Italy). 
Each filter was measured at the room temperature over the whole range from
2000 \AA\ to 1.10 $\mu$m, with readings every 10 \AA.  The light beam had a
diameter of 7 mm and was aimed perpendicularly at the geometrical center of
the filter.  The filter transmissions are given in Table~1 (the full
table is available electronic only), normalized to 1.0 at maximum.  They are
plotted in Figure~1.  We found the difference between the profiles for 2012
and 2014 being negligible, comparable to the small errors of measurements,
indicating a good stability and no (quick) aging of the filters even if in
continuous use at the telescope.  Table~2 present a handy summary of the
basic parameters of the four filters.

\subsection{Stromgren b,y filters}

The Astrodon $b$ and $y$ filters show excellent suppression of residual
trasmittance away from the main profile, with the exception for the $y$
filter of a UV-leak centered at 3275 \AA\, with FWHM=40 \AA\ and 4\% peak
transmission.  Given the low atmospheric transmittance, low CCD sensitivity,
and absence of prominent emission lines at these wavelengths, such a leak
has no practical consequences on photometry of novae.  The Astrodon $b$ and
$y$ filters have effective wavelengths and widths at half maximum (cf. 
Table 2) closely similar to the original Stromgren (1956) interference
filters.  The shape of the transmittance profile is however quite different:
truly flat-topped for Astrodon filters with no significant wings, Gaussian-like
with extended wings for the original Stromgren interference filters.  Such
long wings mean that during the advanced nebular phases, the amount of
[OIII] emission leaking into the $b$ band profile could easily change the
measurement by tenths of a magnitude.  The boxy profile of Astrodon filters
greatly helps in having clean definition of what is transmitted and what is
rejected by the filter.  In addition, a nice advantage of Astrodon filters
is that their dielectric multi-layer type allows peak transmission
efficiencies close to perfect 100\%, while classical interference filters
usually transmit much less than that, requiring proportionally longer exposure
times.

  \begin{figure}[!Ht]
     \centering
     \includegraphics[width=8cm]{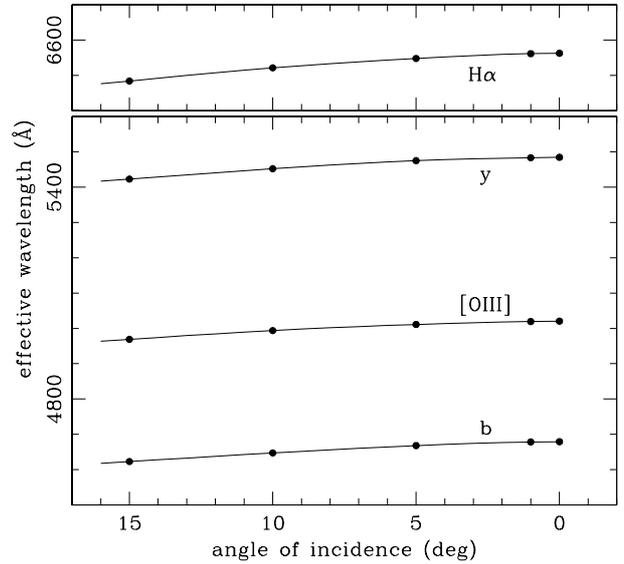}
     \caption{Dependence on angle of incidence of the effective wavelength
     of the Astrodon Stromgren $b$ and $y$ and Baader [OIII] and H$\alpha$
     filters.} 
     \label{fig3}
  \end{figure}

\subsection{Narrow [OIII],H$\alpha$ filters}

The Baader [OIII] and H$\alpha$ filters have high peak-transmission and
Gaussian-like transmission profiles with extended wings.  The [OIII] filter
shows a width at half maximum of 70 \AA\ and it is centered 13~\AA\ to the
red of the rest wavelength of [OIII] 5007.  While this enables the main line
of the doublet to be centered close to the filter peak transmission, the
other line of the multiplet at 4959 \AA\ (whose intensity is about 1/3 of
the 5007 \AA\ one) lies outside the the main transmission profile and on the
wing where the transmission is only 16\% of its peak value, causing a
rejection of about 21\% of the total [OIII] doublet flux.  The H$\alpha$
filter is well centered on the rest wavelength of the emission line and
nicely transmits the whole of it.  

Both Baader [OIII] and H$\alpha$ filters display some leaks away from the
main transmission profile, as shown in Figure~1.  The [OIII] filter has a
near-IR leak at $\lambda$$\geq$10,500 \AA, which has no consequence on Nova
Del 2013 photometry given its blue energy distribution.  On heavily reddened
novae or novae harboring red giants or novae with particularly strong
emission in HeI 10\,830 \AA\ line, the flux recorded through this leak
could however become relevant with respect to that going through the main
transmission profile, even considering the minimal sensitivity of optical
CCDs at such red wavelengths ($\leq$1\%).  This have been clearly shown in
Munari \& Moretti (2012, their Figures~1 and 8) for the 2010 outburst of
nova V407 Cyg, that harbors a very cool Mira.  While the nova was
declining, the flux of the Mira become dominant at the longest wavelengths,
and when the energy distribution became redder than $V$$-$$I_{\rm C}$$\geq$4
mag, the flux going through the similar near-IR leak of old Astrodon $B$
filters become dominant compared to that going through the filter main
profile, resulting in an artificially brighter nova in the $B$-band.  That
near-IR leak has been corrected, and new Astrodon $B$ filters now
offer clean band measurements.

The H$\alpha$ filter has no near-IR leak, but it does not accurately
suppress the transmission away from the main profile, as Figure~1 shows. 
Over the 7000-9000~\AA\ range its average transmission is 0.17 \%.  Again,
this has no practical consequence on Nova Del 2013 photometry given its blue
energy distribution, the lack of significantly strong emission lines in that
wavelength region and the huge intensity of H$\alpha$ emission line.  To
check this, we have integrated the flux through the H$\alpha$ filter profile
on our very extensive set of accurately fluxed spectra of Nova Del 2013 that
cover its entire evolution, and repeated the computation after forcing to
0.00 the filter transmission everywhere outside the main profile.  The
maximum difference between the two sets of measurements amounts to 0.022
mag, for the day of maximum brightness (and lowest equivalent width of
H$\alpha$ emission line).  By the third day past maximum, the difference has
dropped below 0.010 mag, and disappeared below 0.001 mag twenty days past
maximum.  On heavily reddened novae or novae harbouring red giants or novae
with very strong CaII triplet and OI 7772, 8446 \AA\ emission lines, the
story would probably be different and requires checking on a case by case
basis.

\subsection{Depedence on the angle of illumination}

Inserting interference filters in the converging light beam of a telescopes
requires careful evaluation of the effects on the transmission profile.

The wavelength interval transmitted by interference filters shifts toward
the blue when the illuminating beam is tilted away from normal incidence
(90$^\circ$).  We have measured the trasmission profile of all four program
filters for increasing deviations from normal incidence (0$^\circ$,
1$^\circ$, 5$^\circ$, 10$^\circ$ and 15$^\circ$ deg deviations).  These
profiles are plotted in Figure~2 and the corresponding shift in the
effective wavelength is plotted in Figure~3, where the interpolating lines
are drawn from the following parabolic fits:

\noindent
\begin{eqnarray}
\lambda_{\rm eff}({\rm H\alpha})  = 6563.0&    - 1.93{\rm\,\theta}     - 0.223{\rm\,\theta}^2& \\
\lambda_{\rm eff}({\rm y})        = 5484.5&    - 1.18{\rm\,\theta}     - 0.198{\rm\,\theta}^2& \\
\lambda_{\rm eff}({\rm [OIII]})   = 5020.5&    - 1.12{\rm\,\theta}     - 0.154{\rm\,\theta}^2& \\
\lambda_{\rm eff}({\rm b})        = 4679.9&    - 1.85{\rm\,\theta}     - 0.131{\rm\,\theta}^2& 
\end{eqnarray}
where $\lambda$ is in \AA\ and $\theta$ is the deviation (in degrees) from
normal incidence.  Also the width of the profile and peak transmission
change with the deviation, but these are minor effects compare to the
shift in wavelength.

The dependence on the angle of incidence has some relevant consequences. 

A controlled tilt of the filter placed in a {\it parallel} beam can be used to recenter
the passband.  To center the peak of [OIII] filter transmission on the rest
wavelength of [OIII] 5007 line, a tilt of 6$^\circ$.5 should be introduced. 
To center on the [OIII] 5007 line at the barycentric velocities of M33
(Triangulum) and M31 (Andromeda) galaxies, tilts of 7$^\circ$.4 and
8$^\circ$.0 should be applied, respectively.  A similar re-centering for the
H$\alpha$ filter would require tilts of 1$^\circ$.8 and 2$^\circ$.7 for M33
and M31, respectively.

Our observations of Nova Del 2013 were carried out at a f/5.3 Newtonian
focus, by placing the filters in the {\it converging} beam.  This has an angular
aperture of 12$^\circ$.0 (11$^\circ$.67 from the f/number and 0.33 from the
field of view on the sky), fully covered by the oversized filters that do
not introduce spatial vignetting. The mean of the angular deviation from
normal for the light incident on the filters, weighted according to the 
area of the entrance pupil (thus taking into account the obstruction by the
secondary mirror), is
\begin{equation} 
<\theta(f/5.3)>=3^\circ.9
\end{equation}
This, and not 0$^\circ$, is the effective deviation for a target placed 
at the center of the imaging field. The filter parameters for the Newtonian 
f/5.3 focal ratio are listed in Table~2.

\section{Observations}

The photometric observations of Nova Del 2013 were carried out with ANS
Collaboration telescope N.  125 located in Bastia (Ravenna, Italy).  It is a
0.42-m f/5.3 Newton telescope feeding light to a Moravian G2-1600 CCD
camera, equipped with Kodak KAF-1603ME chip, 1536$\times$1024 array, 9
$\mu$m pixels $\equiv$ 0.82$^{\prime\prime}$/pix plate scale.  Technical
details and operational procedures of the ANS Collaboration network of
telescopes running since 2005, are presented by Munari et al.  (2012a). 
Detailed analysis of the photometric performances and measurement of actual
transmission profiles for the photometric filter sets in use with ANS
telescopes is presented by Munari \& Moretti (2012).  All measurements on
Nova Del 2013 were performed with aperture photometry, the long focal length
of the telescope and the sparse surrounding field not requiring the use of
PSF fitting.

  \begin{figure*}[!Ht]
     \centering
     \includegraphics[width=15.2cm]{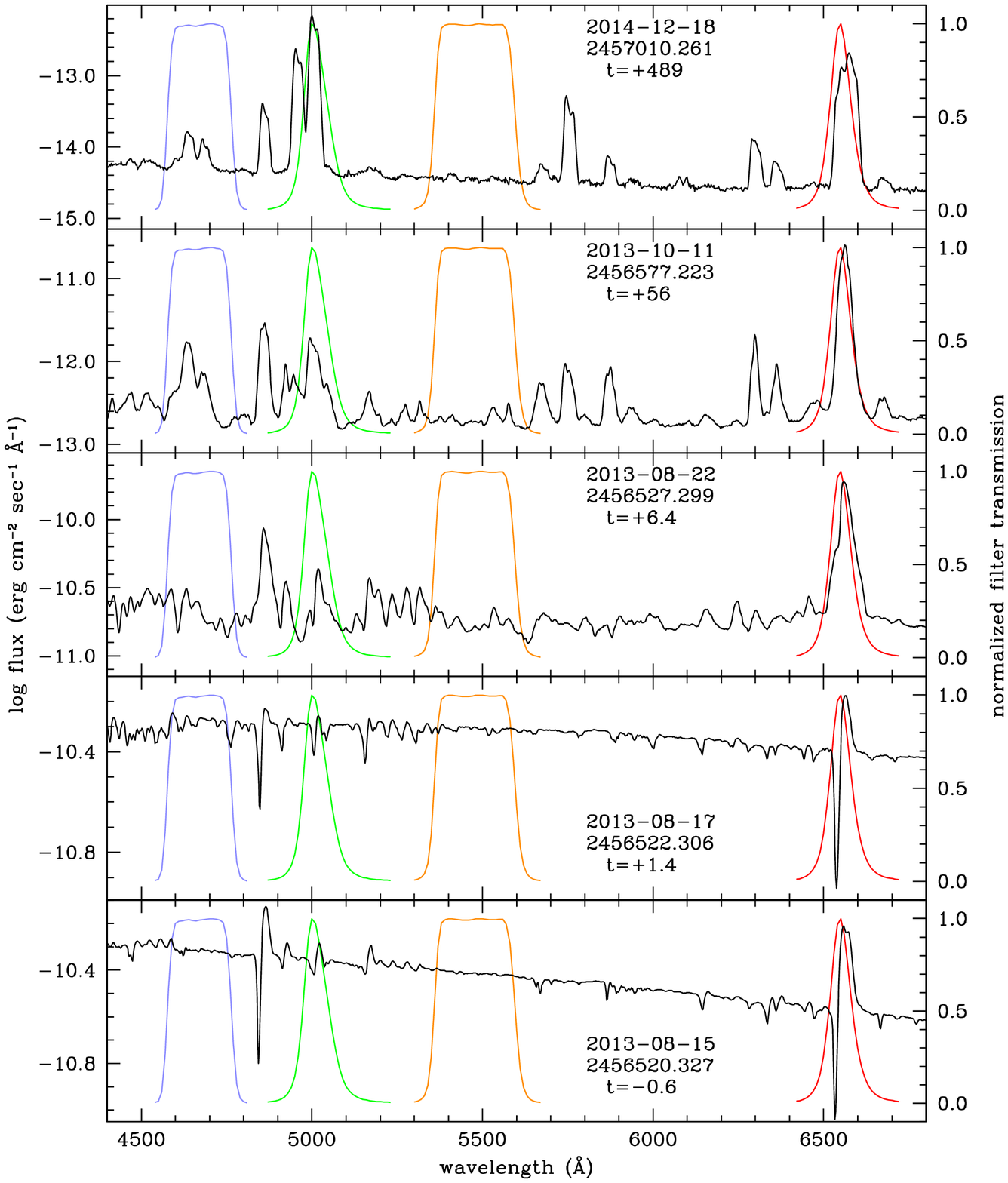}
     \caption{Comparison of the transmission profiles with the spectral
     appearance of Nova Del 2013 at four key dates (time counted from
     maximum brightness on HJD=2456520.9). Note the linear scale
     at right for the filter transmission, and the log scale for the spectra
     at left. From bottom to top panels: spectrum at nova maximum
     brightness, at maximum strength of FeII recombination lines, during the
     "NIII flare" and first emergence of [OIII] lines and, finally, deep
     into the nebular phase.}
     \label{fig4}
  \end{figure*}

  \begin{figure*}[!Ht]
     \centering
     \includegraphics[width=15.2cm]{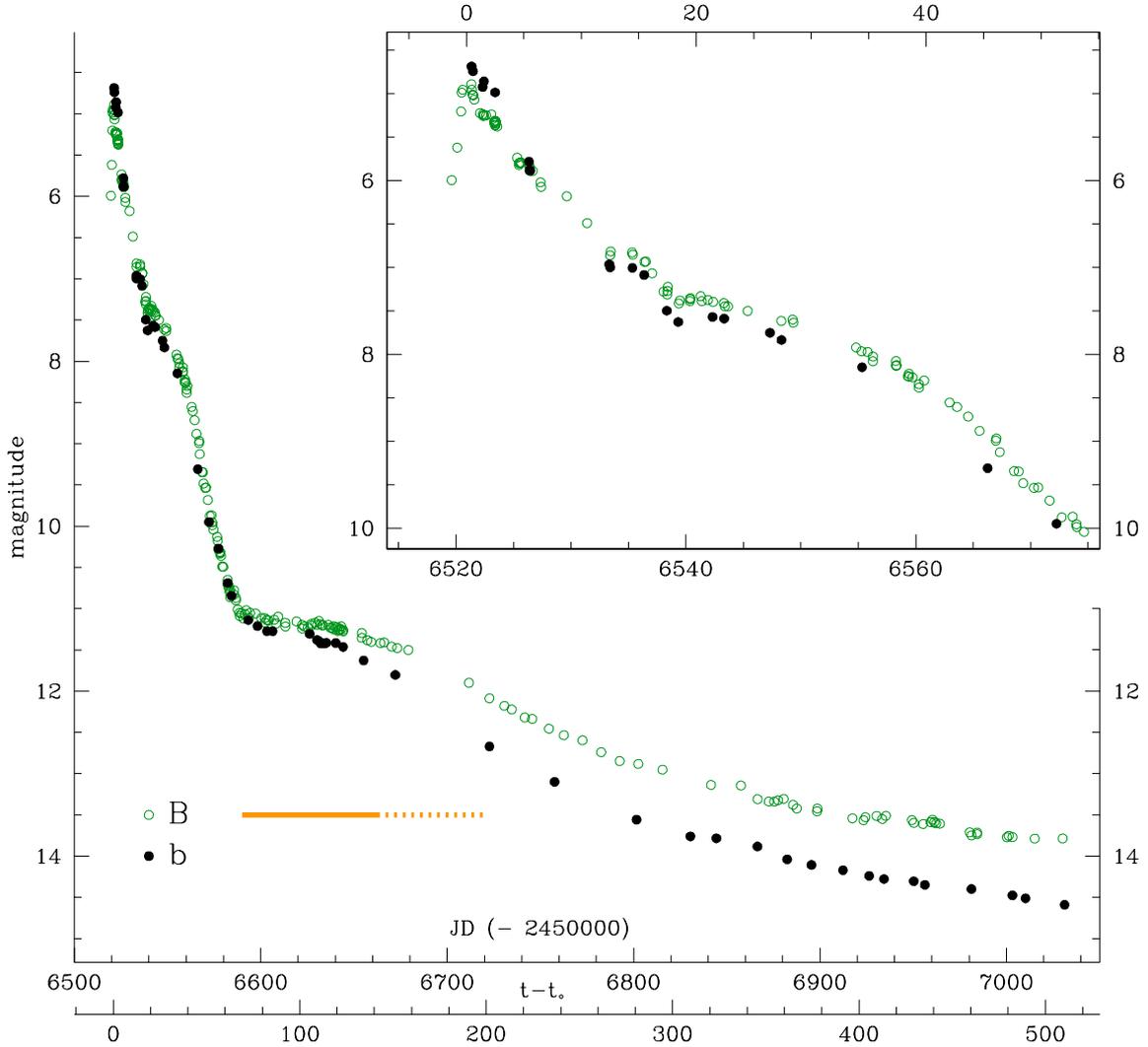}
     \caption{Stromgren $b$-band photometric evolution of Nova Del 2013. 
     The parallel evolution in Johnson's $B$ band is plotted for reference. 
     The insert highlights the evolution around maximum brightness. The
     thick horizontal line marks the period of super-soft X-ray emission
     recorded by Swift satellite, and the dotted part the time interval the
     satellite could not be aimed to Nova Del 2013. When the satellite
     resumed the observations of the nova, the super-soft emission was
     already gone, meaning that the switch off occurred sometime along the
     dotted part of the line.}
     \label{fig5}
  \end{figure*}

  \begin{figure*}[!Ht]
     \centering
     \includegraphics[width=15.2cm]{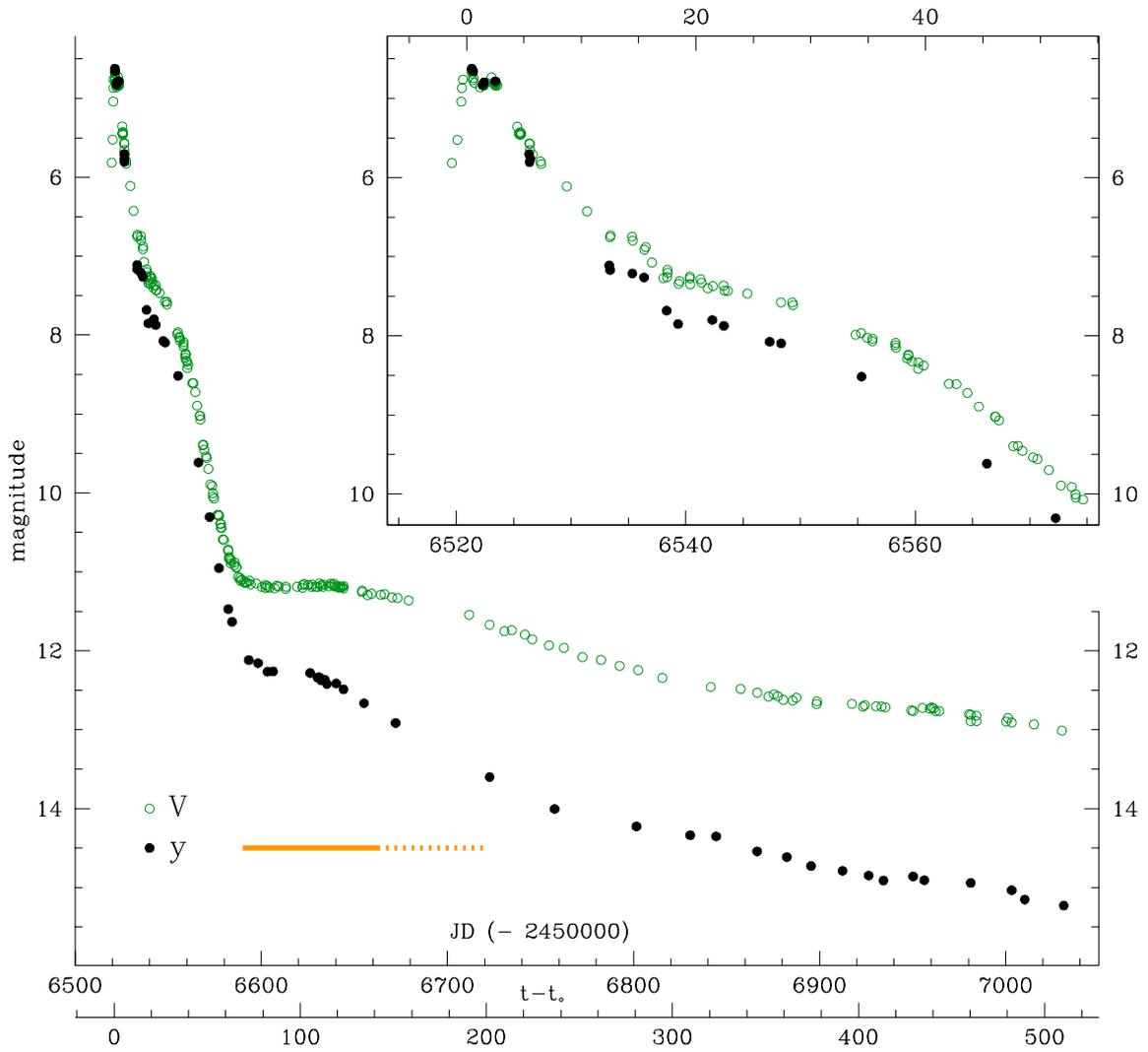}
     \caption{Stromgren $y$-band photometric evolution of Nova Del 2013.
     The parallel evolution in Johnson's $V$ band is plotted for reference.
     The insert highlights the evolution around maximum brightness.
     See Figure~5 for the meaning of the horizontal thick/dotted line.} 
     \label{fig6}
  \end{figure*}

  \begin{figure*}[!Ht]
     \centering
     \includegraphics[width=15.2cm]{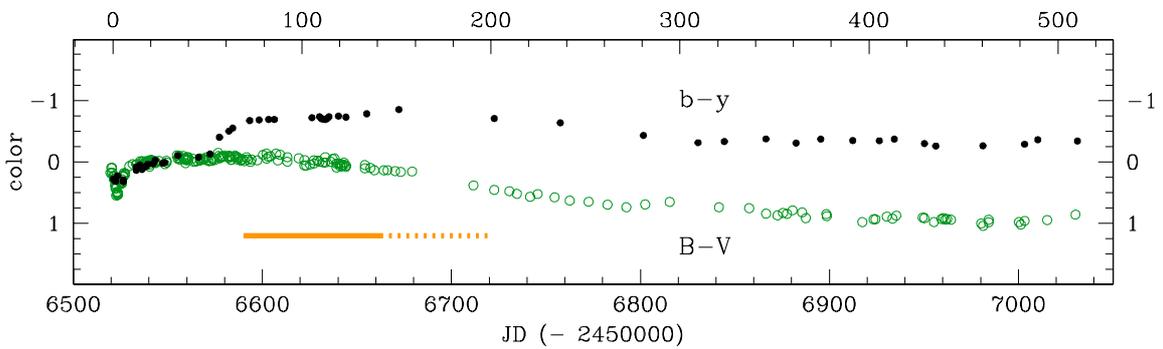}
     \caption{Stromgren $b-y$ photometric color evolution of Nova Del 2013.
     The parallel evolution in Johnson's $B-V$ color is plotted for
     reference. See Figure~5 for the meaning of the horizontal thick/dotted
     line.}
     \label{fig7}
  \end{figure*}

  \begin{figure*}[!Ht]
     \centering
     \includegraphics[width=15.2cm]{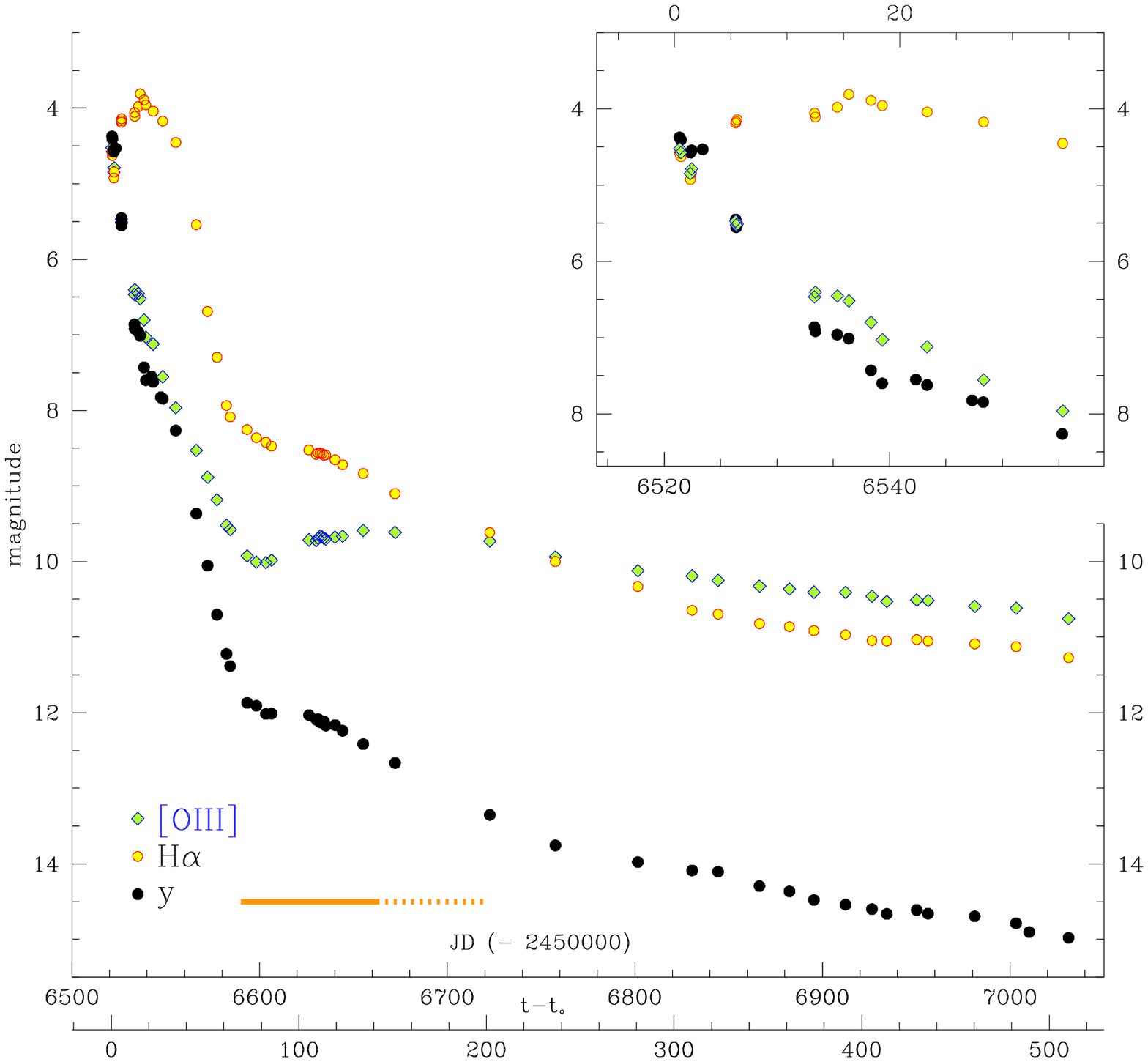}
     \caption{Photometric evolution of Nova Del 2013 as recorded through the
     narrow-band filters centered around [OIII] and H$\alpha$.  The parallel
     evolution in Stromgren $y$-band is re-plotted from Figure~6 for
     comparison.  The insert highlights the evolution around maximum
     brightness. See Figure~5 for the meaning of the horizontal thick/dotted
     line.}
     \label{fig8}
  \end{figure*}

At nova maximum brightness, the exposure times were 25$\times$10 sec in
[OIII], 25$\times$2 sec in H$\alpha$, 30$\times$3 sec in $b$, and
30$\times$2 sec in $y$.  They grew with fading of the nova, reaching for the
latest observations 12$\times$120 sec in [OIII], 12$\times$90 sec in
H$\alpha$, 12$\times$240 sec in $b$, and 12$\times$180 sec in $y$.  A
few-pix wide random dithering was introduced between individual exposures
(each filling about half of the full well capacity), with the observed field
centered in a similar manner on the chip throughout the whole observing
campaign.  This ensured that any issue related to possible outer-field
vignetting or differential illumination by the shutter (of the rotating
blade type) even at the shortest exposures, were accurately cancelled out
throughout the observing campaign.  All photometric measurements were
carefully tied to a local photometric sequence extracted from the APASS
survey (Henden et al.  2012, Henden and Munari 2014, Munari et al.  2014),
covering the whole sky down to $V$=17 in five bands, Landolt $B$,$V$ and
Sloan $g^\prime$,$r^\prime$,$i^\prime$.  Stromgren $b$, $y$ magnitudes of 18
field stars around Nova Del 2013 were derived via the transformation
equations from APASS data published by Munari (2012).  The corresponding
H$\alpha$ and [OIII] magnitudes for these 18 field stars were derived by a
two step procedure.  The magnitudes were first calculated by m(H$\alpha$)=$r^\prime$ and
m([OIII])=($b$+$y$)/2, and further checked that these instrumental magnitudes for
H$\alpha$ and [OIII] images were on a normal Pogson scale.  Secondly, to fix
the zero point, we used the dense spectrophotometric monitoring of Nova Del
2013 that we obtained with the Asiago 1.22m telescope + B\&C spectrograph,
which will be studied elsewhere.  On the absolutely fluxed spectra obtained
at nova maximum brightness, and using the filter transmission profiles for
f/5.3 we computed the [OIII] and H$\alpha$ magnitudes, and repeated the
computation after removing from the spectra the weak emission and
absorption lines present within the filter transmission range.  The
difference between the two readings fixed the zero point of the
transformation scale on the Vega scale.

The same 18 comparison stars were used throughout the whole observing
campaign (and in the process continuously checked for absence of intrinsic
variability).  When the nova was at maximum brightness, they were fainter
than the nova and thus required co-adding (after astrometric registration)
many individual frames (individually calibrated for bias, dark and flat
frames) to obtain a high S/N on the comparison stars, and the reverse toward
the end of the campaign when the nova turned fainter than the comparison
stars.  Even if it was of minimal impact given the narrowness of the filter band
pass, all observations were corrected for instantaneous color equations.

The resulting $b$, $y$, [OIII] and H$\alpha$ photometry of Nova Del 2013 is
given in Tables~2 and 3. The total error budget is quoted for each
measurement, which quadratically includes the Poissonian contributions of the
variable and the comparison stars and the uncertainty in the transformation
to the standard system as defined by the 18 local photometric standards. The
$b$$-$$y$ color is derived directly during the reduction and not computed as
the difference between the corresponding magnitudes.

   \begin{table}[!Ht]
     \caption{Our Stromgren $b$,$y$ photometry of Nova Del 2013.}
      \centering
      \includegraphics[height=170mm]{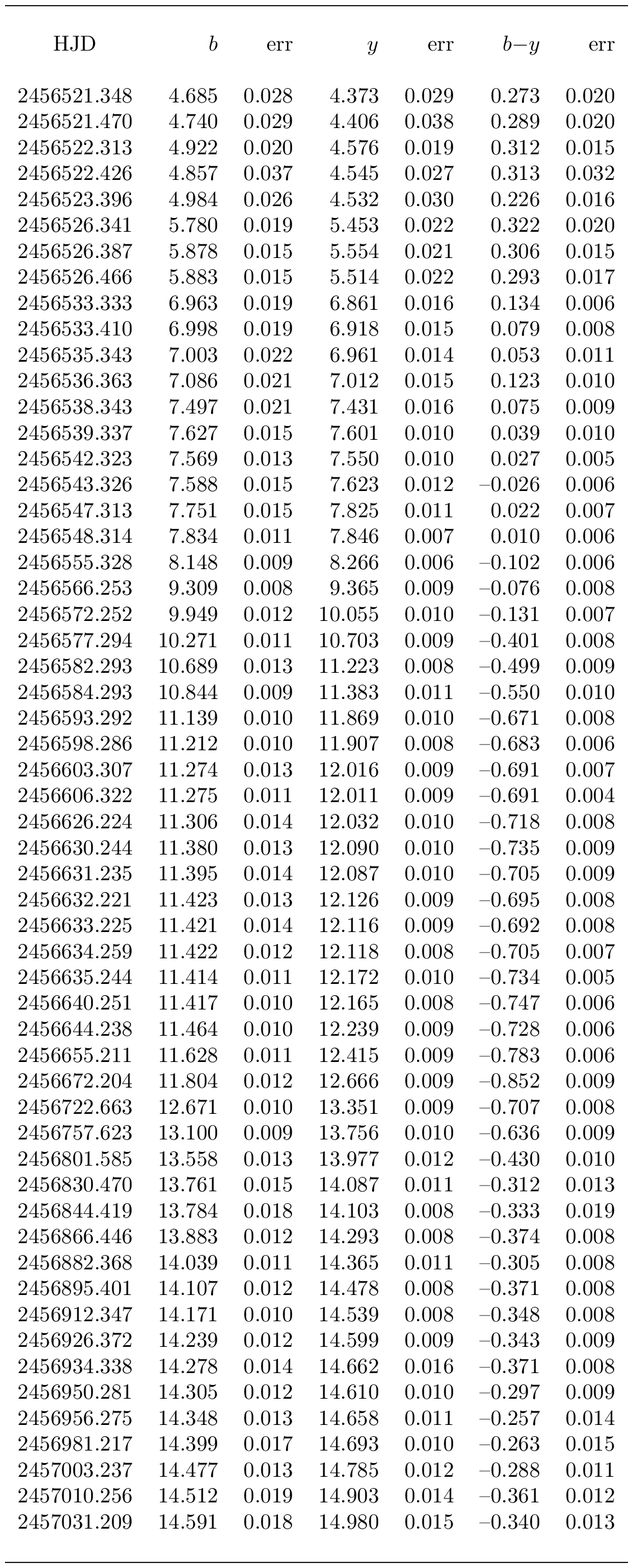}
      \label{tab3}
  \end{table}

   \begin{table}[!Ht]
     \caption{Our [OIII], H$\alpha$ photometry of Nova Del 2013.}
      \centering
      \includegraphics[height=170mm]{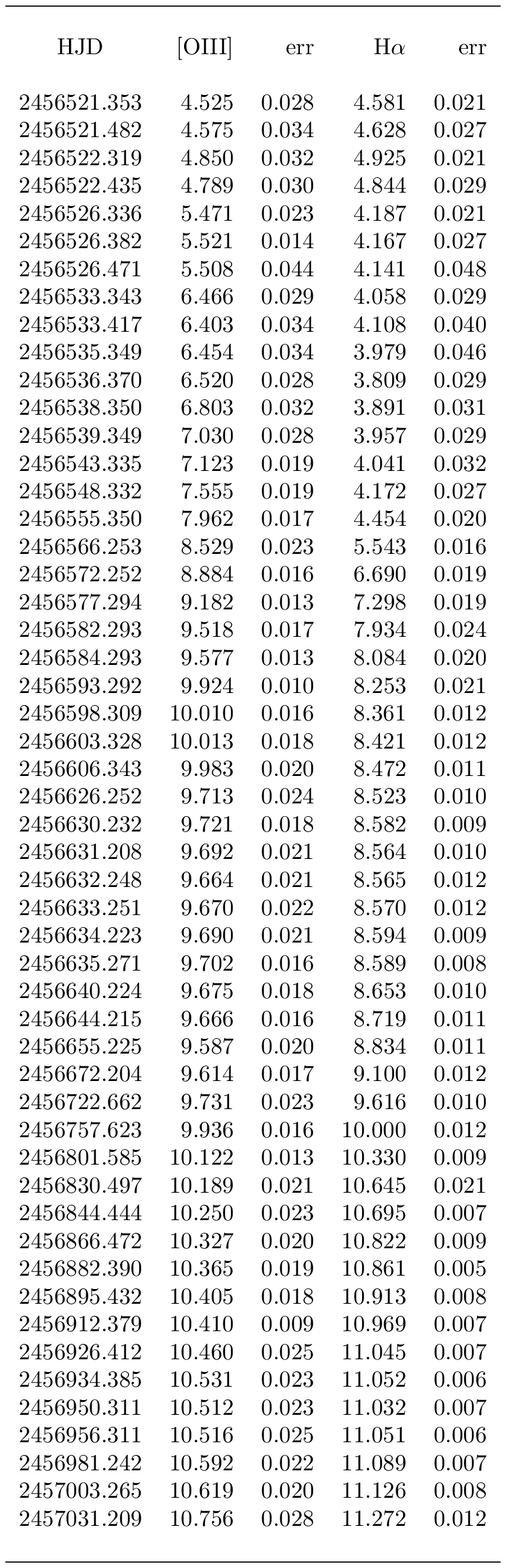}
      \label{tab4}
  \end{table}

\section{Results}

The transmission of the $b$, $y$, [OIII] and H$\alpha$ filters is
overplotted in Figure~4 to the spectral evolution of Nova Del 2013 at some
key epochs.  The photometric evolution in $b$ and $y$ bands is compared to
that in $B$ and $V$ bands in Figures 5 and 6, and the lightcurves in [OIII]
and H$\alpha$ filters are plotted in Figure~8.  The $b$$-$$y$ color evolution
is compared to $B$$-$$V$ in Figure~7, and Figure~9 focuses on the portion of
the lightcurves that cover the period during which super-soft X-ray emission
was recorded from Nova Del 2013.

\subsection{Stromgren bands}

As illustrated by Figure~4, the Stromgren $y$ filter remains remarkably
clear of significant emission lines through the whole photometric evolution,
especially around maximum brightness and advanced decline, confirming that
$y$ band is an excellent tracer of the continuum emission of novae.  At the
time of peak intensity of FeII, a few weak lines from multiplets 48 (5362.9
\AA) and 49 (5425.3 and 5477.2 \AA) emit within the filter passband, and
during early nebular phases weak emission by NII multiplet 63 and the
auroral [NII] 5577 \AA\ lines contribute a little to the recorded flux. 
Around maximum, $V$ and $y$ lightcurves essentially coincide (including the
$\sim$2 day plateau described by Munari et al.  2013d), but as soon as
emission lines emerge with a significant flux a few days past maximum, the
two lightcurves start to diverge with the $V$ band systematically brighter
than $y$ because of the large flux contributed by H$\beta$ and FeII
multiplet 42.  As long as H$\beta$ and FeII 42 are the dominant emission
lines within the $V$ passband, the $V$ and $y$ lightcurves evolve in
parallel with a nearly constant difference in magnitude.  When the ejecta
finally turn optically thin (around day +70), and super-soft X-ray emission
from the central WD begins to be observed, the divergence of $V$ and $y$
lightcurves resumes and continues over the rest of the lightcurve.

When the ejecta turn optically thin (around day +70), the rate of
photometric decline flattens out considerably at all bands.  At this time,
the continuum emission observed from the nova is no more dominated by the
pseudo-photosphere receding through the ejecta, but by the recombinations in
the optically thin ejecta.  Initially (days +70 to +100), the recombinations
are nearly balanced by the re-ionization induced by the hard radiation
emanating from the white dwarf (in the super-soft X-rays phase) that now
permeates the ejecta, and the brightness in the $y$ band remains essentially
constant.  Later on (day +110 and onward), the continuing dilution of the
ejecta into the surrounding space brings recombinations eventually out of
balance with re-ionization, and the $y$ band resumes declining.  Such
decline resumption occurs at later times in the $V$ band (around day +155),
because its bandpass intercepts a large fraction of the increasing emission
in the [OIII] 4959, 5007 \AA\ doublet.  The brightness of the nova in the
[OIII] filter keeps {\em increasing} throughout the super-soft phase, as the
result of the ionization from the central star and the continuous expansion
that brings an increasingly large fraction of the ejecta below the critical
density for collisional de-excitation of [OIII].

  \begin{figure*}[!Ht]
     \centering
     \includegraphics[width=15.2cm]{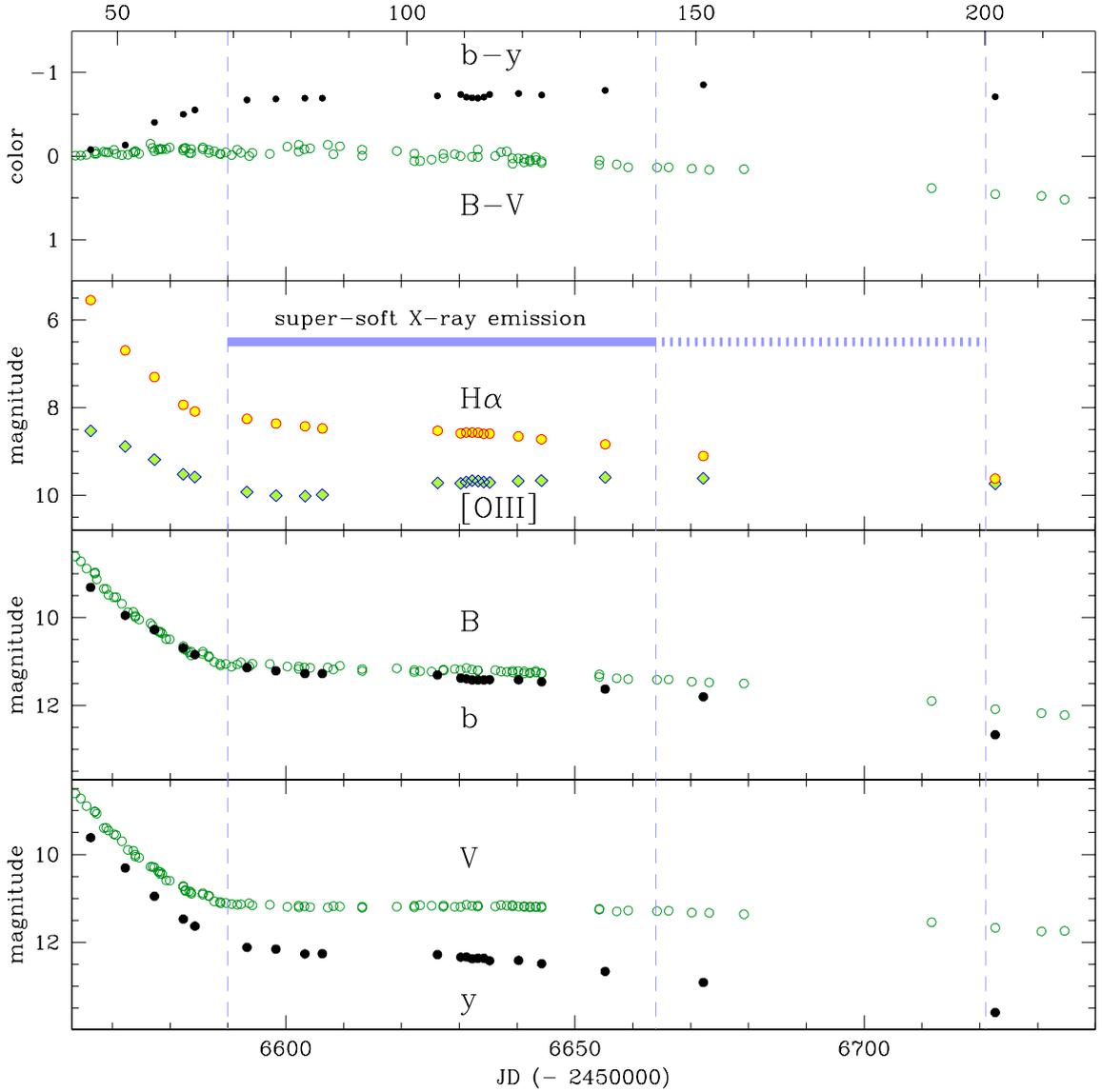}
     \caption{Photometric evolution of Nova Del 2013 around the 
     super-soft X-ray phase. The panels are zooms from Figures
     5$-$8. See Figure~5 for the meaning of the horizontal thick/dotted
     line on the [OIII], H$\alpha$ panel.}
     \label{fig9}
  \end{figure*}

The Stromgren $b$ band is far more affected by emission lines than the $y$
band (cf Figure~4), and it evolves much closer to $B$ band (cf Figure~5)
than $y$ compared to $V$.  In particular, no plateau soon after optical
maximum is observed in $b$, and it is similarly missing in $B$.  Around maximum
brightness, Nova Del 2013 is brighter in $b$ than $B$ because of the greater
density of deep aborption lines in $B$ compared to $b$, especially in the
bluer part of the $B$ passband.  As the Balmer and FeII emission lines
begin to reinforce and the absorption weakens, $b$ and $B$ are first equal
and then $b$ turns slightly fainter than $B$ because no Balmer and only a
few weak lines from FeII multiplets 37, 38, 43 and 54 emit within the $b$
passband.  As for the $y$ band with respect to $V$, the small offset between
$b$ and $B$ remains stable (days +12 to +45 in Figure 5) as long as the
spectra are dominated by Balmer and FeII emission lines.  Before novae enter
the nebular phase, they usually experience the {\em NIII flaring} phase
(McLaughlin 1960), during which the 4640 \AA\ blend of fluorescent excited
NIII lines greatly increase in intensity, surpassing that of H$\gamma$. 
Because the NIII 4640 \AA\ blend sits entirely within the $b$ passband, at
the peak of the NIII flaring (days +60 to +65) the brightness of Nova Del
2013 in $b$ increases and equals
the $B$ magnitude.  Soon after, at the time of the
emergence of super-soft X-ray emission and simultaneous flaring of [OIII]
and decline of NIII 4640 \AA\ blend, the brightness in $b$ and $B$ bands
starts to depart and continue to diverge for the rest of the decline toward
quiescence.  This divergence is however much smaller than observed between
$y$ and $V$ bands (compare Figure 5 and 6), with the brightness in the $b$
band sustained by HeII 4686 \AA\ emission line in addition to that of NIII
4640 \AA\ blend during the advanced decline.

\subsection{[OIII] and H$\alpha$ bands} 

The photometric evolution of Nova Del 2013 in the [OIII] and H$\alpha$
narrow-bands, as depicted in Figure~8, is particularly interesting. 

In broad band photometry, soon after maximum brightness, Nova Del 2013 went
through a brief plateau which lasted for $\sim$2 days in $V$, $\sim$3 in
$R_{\rm C}$ and $\sim$4 in $I_{\rm C}$ (Munari et al.  2013d).  Such a
plateau has no counterpart in [OIII] and H$\alpha$ bands, where the decline
started immediately after maximum was reached (see the first two observing
dates on the upper right panel of Figure~8).

Around nova maximum brightness, the flux recorded through the [OIII] band is
essentially that of the continuum, because the strong FeII multiplet 42 line
at 5018 \AA\ presents a marked P-Cyg profile (cf.  Figure~4), where FeII
absorption and emission balance each other.  As soon as the P-Cyg absorption
component vanishes, the nova brightness in the [OIII] filter rises above
that of the Stromgren $y$ band recording the adjacent continuum, and remains
brighter as long as the FeII multiplet 42 stays in emission.  With the change
to optically thin conditions of the ejecta and the spreading ionization
caused by the super-soft X-ray input, the [OIII] 5007 \AA\ emission
line rapidly grows in intensity.  This causes the brightness in the [OIII]
band to slow and then to stop the decline (initial 20 days of the super-soft
phase), followed by an actual and long lasting {\em brightening} (cf. 
Table 4 and Figure~9).  At the end of the super-soft X-ray phase (130 days
past it begun), Nova Del 2013 is still brighter by 0.2 mag in the [OIII] band than
before it started.  If confirmed in other novae, this increase and plateau
phase in [OIII] band brightness could be used to trace the period during
which super-soft X-ray emission is present.  Following the end of
the super-soft X-ray phase, the [OIII] band brightness begun to decline, at
a very slow rate, slower than in the H$\alpha$ band.  At the time of the our
last observation, on day +510, in the [OIII] band Nova Del 2013 was 4.22 mag
brighter than in $y$ and 2.27 mag brighter than in $V$.

As for [OIII], the flux recorded around maximum through the H$\alpha$ band
is essentially that of the continuum, because the strong H$\alpha$ line at
6563 \AA\ presents a marked P-Cyg profile (cf.  Figure~4), where again
absorption and emission balance each other.  After maximum brightness, the
pseudo-photosphere begins to contract inward through the ejecta and to rise
in $T_{\rm eff}$, so that a growing fraction of gas in front of the
pseudo-photosphere is ionized, causing a brightening in H$\alpha$ band,
reaching maximum on day +16 for Nova Del 2013, and remaining on a plateau
for about 35 days past the maximum brightness observed in the $V$ band (cf
Figure~8).  At the time when the plateau ended, Nova Del 2013 was in the
H$\alpha$ band 3.6 mag brighter than in $V$, 3.8 mag brighter than in $y$,
and 3.5 mag brighter than in the [OIII] band.  After the plateau phase, the
brightness in the H$\alpha$ band began to decline very rapidly (0.13 mag
day$^{-1}$), at a speed similar to that of the continuum (as traced by the
Stromgren $y$ band) was declining (0.12 mag day$^{-1}$) during the H$\alpha$
plateau (cf.  Figure~8).  The post-plateau drop in H$\alpha$ brightness
coincides precisely with the first appearance of the [OIII] 5007 emission line
and the NIII 4640 \AA\ flaring.  The rapid drop in the H$\alpha$ band was
halted by the start of the super-soft X-ray phase, during which the decline
in H$\alpha$ band proceeded at a much slower rate (0.010 mag day$^{-1}$). 
At the end of the super-soft X-ray phase, the [NII] 6548, 6584 \AA\ doublet
was already contributing significantly to the brightness of Nova Del 2013 in
the H$\alpha$ band, and this contribution increased with time, so that by
June/July 2014 nearly all the flux recorded in the H$\alpha$ band was coming
from the [NII] doublet.  Once this happened, the brightness decline of Nova
Del 2013 in the H$\alpha$ band became slower than that of the [OIII]
band, and the decline in brightness proceeded in parallel at the same
0.0027 mag day$^{-1}$ for both the [OIII] and H$\alpha$ bands.

The extreme smoothness of both the H$\alpha$ and [OIII] lightcurves argues
for absence of large and abrupt discontinuities in the ejecta of Nova Del
2013. Should they exist, glitches in the lightcurves would have appeared
when the ionization and/or recombination fronts overtook them.

\subsection{Optical photometry during the phase of large variability in
the super-soft X-ray emission}

During the early portion of the super-soft X-ray emission phase of novae,
the super-soft flux is frequently seen to vary by huge amounts over short
time-scales (Ness et al.  2007, Osborne et al.  2011, Osborne et al.  2013a,
Schwarz et al.  2011).  Beardmore et al.  (2013) reported for Nova Del 2013
variations up to 5 magnitudes (i.e.  100 times) from one day to the next in
the super-soft X-ray flux recorded during the two-weeks rise toward maximum
X-ray brightness, which covers the period of time from day +69 to day +86
(cf.  Figure~9).  This effectively corresponds to turning on/off the
recorded X-ray emission.  The origin of these large amplitude variations
(LAVs) in the super-soft X-ray flux is not certain, but it has been
associated to clumpy ejecta passing through the line of sight or white dwarf
photospheric temperature variations.

Our photometry in Figure~9 obtained during the LAVs period of Nova Del 2013
is particularly accurate and helps to constrain the origin of LAVs.  The
total error budget of our photometric measurements is much smaller than dot
dimensions in Figure~9, as also confirmed by the residuals from low-order
polynomial fits to the lightcurves which are of the order of 0.010 mag.  A
most relevant feature of the photometry in Figure~9 is that there is not
even the lightest trace of short term optical variability superimposed to
the underlying smooth trend during the LAVs period.  This is equally true
for narrow-band [OIII] and H$\alpha$, for Stromgren $b$ and $y$ and for
Johnson-Cousins $B$ and $V$ bands.  So no matter if the measurement refers
only to continuum radiation, where all the ejecta contribute to the recorded
flux (Stromgren $y$), or just to a single emission line, where radiation is
received only from a single layer of the ejecta ([OIII] and H$\alpha$), the
X-ray LAVs do not extend into the optical.

The presence of LAVs during the time interval from day +69 to day +86 is
simultaneous with the initial growth in intensity of the [OIII] 4959, 5007
\AA\ lines in the optical spectra.  Standard photo-ionization modeling with
CLOUDY shows that in such conditions the [OIII] emission originates from a
confined region of the ejecta, extending radially for 10\%$-$20\% of the
total thickness of the ejecta (e.g.  Munari et al.  2008).  The critical
electron density for the collisional de-excitation of these lines is
$\sim$6.8$\times$10$^5$ cm$^{-3}$ (Osterbrock and Ferland 2006), which is
therefore the electron density of the ejecta or at least of the region
producing the [OIII] lines when they turn visible.  Such an electron density
corresponds to a recombination time scale of one week (Ferland 2003).  The
fact that LAVs seen in Nova Del 2013 did not impact the brightness of the
ejecta by an amount larger than the negligible measurement errors (of the
order of 1\%), indicates that LAVs are not related to global changes
affecting the white dwarf as a whole, like the proposed variability of the
surface temperature.  Should this be the reason for observed LAVs, it would
imply a parallel large variability of the input of ionizing radiation to the
ejecta, whose balance with the recombinations ultimately drives the observed
optical brightness of the ejecta.  The LAVs time scale observed in Nova Del
2013 is of the order of one day, while the recombination time scale is a
week.  The two are sufficiently similar that one would expect to observe at
optical wavelengths a counterpart of LAVs, which is not seen.  The only feasible
possibility is that LAVs affect only the x-ray emission directed toward the
Sun, as with clumpy ejecta passing through the line of sight to us, leaving
unchanged the hard radiation input to the ejecta integrated over the 4$\pi$
sky as seen from the white dwarf.

\section{Acknowledgements}

We would like to thank Arne A. Henden and Elena Mason for careful reading
of the manuscript and detailed comments.

\end{document}